\documentclass[a4paper,12pt]{article}
\usepackage[brazil]{babel}
\usepackage{amsfonts}
\usepackage{graphicx}
\usepackage{amsmath}

\begin{document}

\begin{center}
{\huge Esferas de G\'{a}s em Estado de Equil\'{\i}brio sob A\c{c}\~{a}o da
Gravita\c{c}\~{a}o Pr\'{o}pria}

\textbf{Andrei Smirnov\footnote{%
smirnov@ufs.br; smirnov.globe@gmail.com}, Ricardo Max Menezes Oliveira%
\footnote{%
ricardoxam@gmail.com}}

\textit{Universdade Federal de Sergipe}

Resumo
\end{center}

{\small Nesse pequeno trabalho discutimos estados de equil\'{\i}brio de
estrealas, consideradas como um corpo de g\'{a}s sujeito a a\c{c}\~{a}o da
for\c{c}a gravitacional pr\'{o}pria, utilizando um modelo anal\'{\i}tico
simplificado. Usamos uma condi\c{c}\~{a}o para o estado de equil\'{\i}brio
de um corpo de g\'{a}s na forma de uma equa\c{c}\~{a}o diferencial que
relaciona distribui\c{c}\~{a}o de press\~{a}o e densidade de massa no corpo.
Foram discutidas as distribui\c{c}\~{o}es de densidade de massa da forma
constante, potencial, exponencial e gaussiana. Foram obtidas as express\~{o}%
es exatas para distribui\c{c}\~{a}o de massa e press\~{a}o na dire\c{c}\~{a}%
o radial e a press\~{a}o central.}

\textbf{Palavras-chave}: modelo anal\'{\i}tico estrelar, estado de equil%
\'{\i}brio, solu\c{c}\~{o}es exatas.

\begin{center}
{\huge Self-graviting Gas Spheres in Equilibrium State}

Abstract
\end{center}

{\small In the paper we discuss equilibrium states of stars, using a
simplified analytic model. A star is considered as self-graviting body of
gas. We use a condition for the equilibrium state of the body in the form of
a differential equation, which relates the pressure distribution and mass
density in the body. The density distributions of constant, potential,
gaussian, and exponential forms are discussed. Exact expressions for the
distribution of mass and pressure along the radial direction, and the
central pressure were obtained.}


\section{Introdu\c{c}\~{a}o}

De acordo com uma vis\~{a}o moderna uma estrela \'{e} considerada como uma
esfera massiva de g\'{a}s (ou mais exato de plasma), mantida \'{\i}ntegra
pela for\c{c}a de gravita\c{c}\~{a}o pr\'{o}pria. A estrutura interna de
estrelas \'{e} bastante complicada e depende de massa estrelar. Para
estrelas de massa solar (0,5-1,5 de massa solar) da sequ\^{e}ncia principal
da diagrama de Hertzsprung-Russell a estrutura interna inclui: um n\'{u}%
cleo, uma zona de radia\c{c}\~{a}o e uma zona de covec\c{c}\~{a}o. As
estrelas massivas (com massa maior de 1,5 de massa de Sol) possuem n\'{u}%
cleo convectivo acima de qual \'{e} localizada a zona de radia\c{c}\~{a}o
(ou um envelope de radia\c{c}\~{a}o). Nas estrelas de massa menor de 0,5 de
massa de Sol a zona de radia\c{c}\~{a}o \'{e} ausente (diz-se que estrelas
possuem um envelope de convec\c{c}\~{a}o). Nas estrelas da sequ\^{e}ncia
principal a energia est\'{a} sendo gerada pela queima de hidrog\^{e}nio em h%
\'{e}lio atrav\'{e}s de fus\~{a}o nuclear em seu n\'{u}cleo.

Na classifica\c{c}\~{a}o de estrelas fora da sequ\^{e}ncia principal s\~{a}o
distinguidas estrelas de tipos: subgigantes, gigantes, gigantes luminosas,
supergigantes, hipergigantes, sub-an\~{a}s, estrelas retardat\'{a}rias
azuis. Nas estrelas destes tipos a energia \'{e} gerada principalmente pelo
mecanismo de fus\~{a}o nuclear. Al\'{e}m disso s\~{a}o consideradas as
estrelas pr\'{e}-sequ\^{e}ncia principal. A fonte de energia desses objetos 
\'{e} causada somente pela contra\c{c}\~{a}o gravitacional em oposi\c{c}\~{a}%
o \`{a} fus\~{a}o nuclear em estrelas de outros tipos. Estrelas de v\'{a}%
rios tipos possuem caracter\'{\i}sticas espec\'{\i}ficas de estrutura,
mecanismos da produ\c{c}\~{a}o e transporte de energia. Para alguns tipos de
estrelas s\~{a}o constru\'{\i}dos os bons modelos que descrevem
adequadamente as caracter\'{\i}sticas observadas das estrelas, para outros
tipos os modelos est\~{a}o em processo de constru\c{c}\~{a}o e discuss\~{a}o.

O modelo mais simples de estrutura estelar \'{e} a aproxima\c{c}\~{a}o
quase-est\'{a}tica de simetria esf\'{e}rica. Uma discuss\~{a}o desse modelo 
\'{e} apresentada em Refs. \cite{Clayton}, \cite{Hansen}. O modelo assume
que a estrela possui simetria esf\'{e}rica e est\'{a} em estado de equil%
\'{\i}brio hidrost\'{a}tico. O modelo \'{e} baseado em um conjunto de equa%
\c{c}\~{o}es diferenciais ordin\'{a}rias. Duas equa\c{c}\~{o}es descrevem
varia\c{c}\~{a}o de mat\'{e}ria e press\~{a}o na dire\c{c}\~{a}o radial.
Outras equa\c{c}\~{o}es descrevem varia\c{c}\~{a}o da luminosidade e
transporte de energia. Para determina\c{c}\~{a}o da luminosidade, precisa da
caracter\'{\i}stica taxa de produ\c{c}\~{a}o de energia. Esta caracter\'{\i}%
stica \'{e} determinada atrav\'{e}s de um mecanismo principal de transporte
de energia: radia\c{c}\~{a}o ou convec\c{c}\~{a}o. O conjunto das equa\c{c}%
\~{o}es do modelo comp\~{o}e um sistema das equa\c{c}\~{o}es n\~{a}o
lineares, que pode ser resolvido em geral por m\'{e}todos num\'{e}ricos.

De fato nesse modelo as duas equa\c{c}\~{o}es para mat\'{e}ria e press\~{a}o
s\~{a}o desacopladas das outras equa\c{c}\~{o}es, pois elas n\~{a}o cont\'{e}%
m par\^{a}metros como temperatura, taxa de produ\c{c}\~{a}o de energia,
luminosidade, portanto podem ser consideradas separadamente. Impondo uma fun%
\c{c}\~{a}o da distrubui\c{c}\~{a}o de densidade da materia na estrela, \'{e}
possivel obter a distribui\c{c}\~{a}o de press\~{a}o. Adicionando uma equa%
\c{c}\~{a}o de estado para par\^{a}metros termodin\^{a}micos, como por
exemplo a equa\c{c}\~{a}o de estado de um g\'{a}s perfeito, \'{e} poss\'{\i}%
vel calcular a temperatura na estrela. Adicionando ainda uma rela\c{c}\~{a}o
apropriada para taxa de produ\c{c}\~{a}o de energia, \'{e} possivel estimar
a luminosidade da estrela. De tal maneira pode ser constru\'{\i}do um modelo
anal\'{\i}tico. Claro que esse modelo \'{e} simplificado, mas permite fazer
estima\c{c}\~{o}es de algumas caracter\'{\i}sticas principais de estrela,
tais como: densidade central, temperatura central, luminosidade em termos de
raio e massa de estrela. Tal forma de constru\c{c}\~{a}o de modelo anal\'{\i}%
tico foi discutida em Ref. \cite{stein}, onde foi usada a distribui\c{c}\~{a}%
o de densidade na forma linear, chamado modelo linear de estrela.
Entretanto, o uso da densidade na forma linear n\~{a}o parece muito real%
\'{\i}stico.

Nesse pequeno trabalho usamos os modelos anal\'{\i}ticos com v\'{a}rias
distribui\c{c}\~{o}es de densidade com aspectos mais real\'{\i}sticos. Uma
peculiaridade dos modelos propostos no trabalho \'{e} a possibilidade de
resolver as equa\c{c}\~{o}es diferenciais analiticamente. No trabalho
focalizamos na obten\c{c}\~{a}o da distribui\c{c}\~{a}o de press\~{a}o na
dire\c{c}\~{a}o radial e press\~{a}o central, ignorando a determina\c{c}\~{a}%
o de outros par\^{a}metros. Discutimos os modelos com distribui\c{c}\~{o}es
de densidade das formas potencial, exponencial e gaussiana. No in\'{\i}cio
discutimos tamb\'{e}m um modelo com densidade constante para demonstrar
transparentemente os passos realizados nos c\'{a}lculos, e para introduzir
nota\c{c}\~{o}es usadas posteriormente. As distribui\c{c}\~{o}es de massa e
de press\~{a}o obtemos na forma anal\'{\i}tica. Para realizar os c\'{a}%
lculos foram usados v\'{a}rios m\'{e}todos de f\'{\i}sica matem\'{a}tica
dispon\'{\i}veis na literatura, por exemplo \cite{boas}, \cite{ll-t2}, \cite%
{schutz}. Para visualiza\c{c}\~{a}o e compara\c{c}\~{a}o dos resultados
apresentamos tamb\'{e}m os resultados graficamente. Na conclus\~{a}o
discutimos os resultados obtidos e suas aplica\c{c}\~{o}es poss\'{\i}veis.

Um dos aspectos motivacionais deste trabalho \'{e} a sua aplica\c{c}\~{a}o
pedag\'{o}gica. No trabalho demostramos de forma concisa a constru\c{c}\~{a}%
o das equa\c{c}\~{o}es e da condi\c{c}\~{a}o de contorno, que descrevem um
modelo anal\'{\i}tico e simplificado das estrelas. Isto \'{e}, a equa\c{c}%
\~{a}o de equil\'{\i}brio hidrost\'{a}tico e a equa\c{c}\~{a}o de conserva%
\c{c}\~{a}o de massa. Supomos que os modelos discutidos neste trabalho podem
ser utilizados em salas de aula ou como exerc\'{\i}cios em disciplinas
introdut\'{o}rias de astrof\'{\i}sica, disciplinas dedicadas a modelagem de
fen\^{o}menos f\'{\i}sicos, ou at\'{e} mesmo em equa\c{c}\~{o}es
diferenciais ordin\'{a}rias.

\section{Esferas de Equil\'{\i}brio}

\begin{figure}[th]
\begin{center}
\includegraphics[width=12cm]{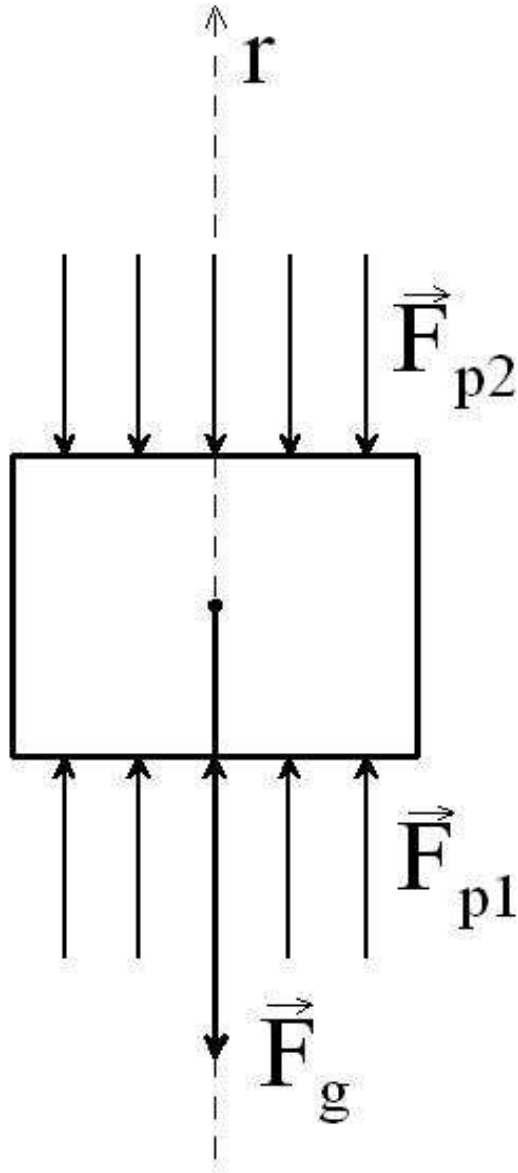}
\end{center}
\caption{Esquema das for\c{c}as externas atuando no fragmento de volume
infinitesimal.}
\label{f2.0s}
\end{figure}

Consideraremos um volume de g\'{a}s esfericamente sim\'{e}trico de raio $R$
em estado de equl\'{\i}brio. A coordenada radial $r$ dentro da esfera varia
como $0\leq r\leq R$. Escolhemos um fragmento de volume dentro da esfera da
forma de um paralelep\'{\i}pedo ret\^{a}ngulo infinitesimal $dV=dr\Delta s$
com base $\Delta s$ e de altura $dr$ (ver Fig. \ref{f2.0s}). No fragmento de
volume atuam as seguintes for\c{c}as externas: a for\c{c}a superficial de
press\~{a}o das camadas adjacentes de g\'{a}s e a for\c{c}a gravitacional
volum\'{e}trica. Para garantir equilibrio do fragmento dentro da esfera de g%
\'{a}s a for\c{c}a gravitacional $\vec{F}_{g}$ deve ser equilibrada pelas for%
\c{c}as de press\~{a}o que atuam na face inferior do fragmento $\vec{F}_{p1}$
e na face superior do fragmento $\vec{F}_{p2}$ (for\c{c}as de press\~{a}o
perperndicular a dire\c{c}\~{a}o radial s\~{a}o equilibradas por causa de
simetria esf\'{e}rica e n\~{a}o s\~{a}o indicadas na figura), portanto:%
\begin{equation}
\vec{F}_{g}+\vec{F}_{p1}+\vec{F}_{p2}=0~,  \label{2.01}
\end{equation}%
onde: 
\begin{equation}
F_{g}=G\frac{M_{r}dm}{r^{2}},\ F_{p1}=p_{1}\Delta s,\ F_{p2}=p_{2}\Delta s~,
\label{2.02}
\end{equation}%
$dm=\rho dV=\rho dr\Delta s$ \'{e} a massa do fragmento infinitesimal
considerado, $\rho $ \'{e} a densidade da subst\^{a}ncia de g\'{a}s, $%
p_{1}=p\left( r\right) $, $p_{2}=p\left( r+dr\right) $ s\~{a}o press\~{o}es
nos pontos indicados, $M_{r}$ \'{e} a massa dentro da esfera de raio $r$%
\begin{equation}
M_{r}=\int_{0}^{r}4\pi \rho \left( r\right) r^{2}dr~.  \label{2.03}
\end{equation}%
Levando em conta os sentidos das for\c{c}as, temos:%
\begin{equation}
-G\frac{M_{r}\rho dr\Delta s}{r^{2}}+p\left( r\right) \Delta s-p\left(
r+dr\right) \Delta s=0~.  \label{2.04}
\end{equation}%
Apresentando a diferen\c{c}a das press\~{o}es nos pontos pr\'{o}ximos:%
\begin{equation}
p\left( r\right) -p\left( r+dr\right) =-\frac{dp}{dr}dr,  \label{2.05}
\end{equation}%
chegamos \`{a} equa\c{c}\~{a}o diferencial ordin\'{a}ria%
\begin{equation}
\frac{dp}{dr}=-G\frac{M_{r}\rho }{r^{2}}~,  \label{2.06}
\end{equation}%
que relaciona a fun\c{c}\~{a}o de distribui\c{c}\~{a}o de densidade da subst%
\^{a}ncia de g\'{a}s $\rho \left( r\right) $ e press\~{a}o $p\left( r\right) 
$. A Eq. (\ref{2.06}) apresenta a equa\c{c}\~{a}o de Euler para um fluido em
estado de equil\'{\i}brio para o modelo considerado. Na superf\'{\i}cie da
esfera a press\~{a}o \'{e} nula. Obtendo assim a condi\c{c}\~{a}o de
contorno no ponto $r=R$:%
\begin{equation}
p\left( R\right) =0~.  \label{2.07}
\end{equation}

Mostraremos v\'{a}rias formas de distrubui\c{c}\~{a}o de densidade da subst%
\^{a}ncia na esfera que faz com que obtenhamos solu\c{c}\~{o}es exatas para
distribui\c{c}\~{a}o da press\~{a}o. 

\subsection{Desidade Constante}

\begin{figure}[th]
\begin{center}
\includegraphics[width=12cm]{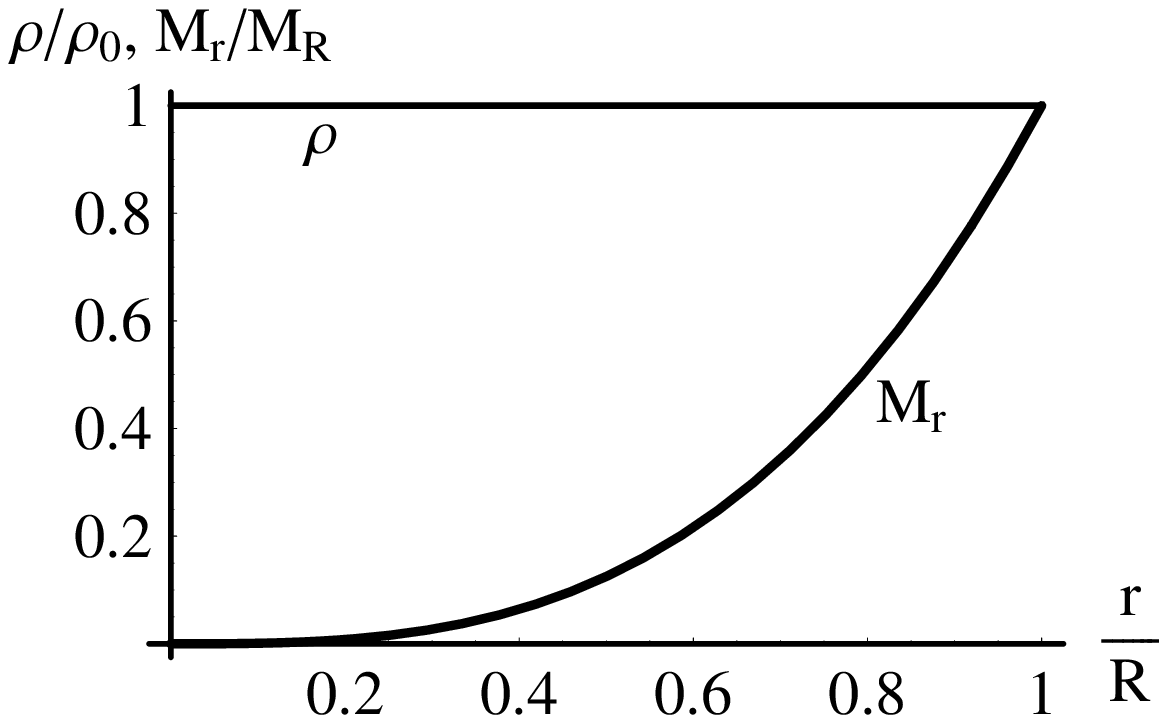} %
\includegraphics[width=12cm]{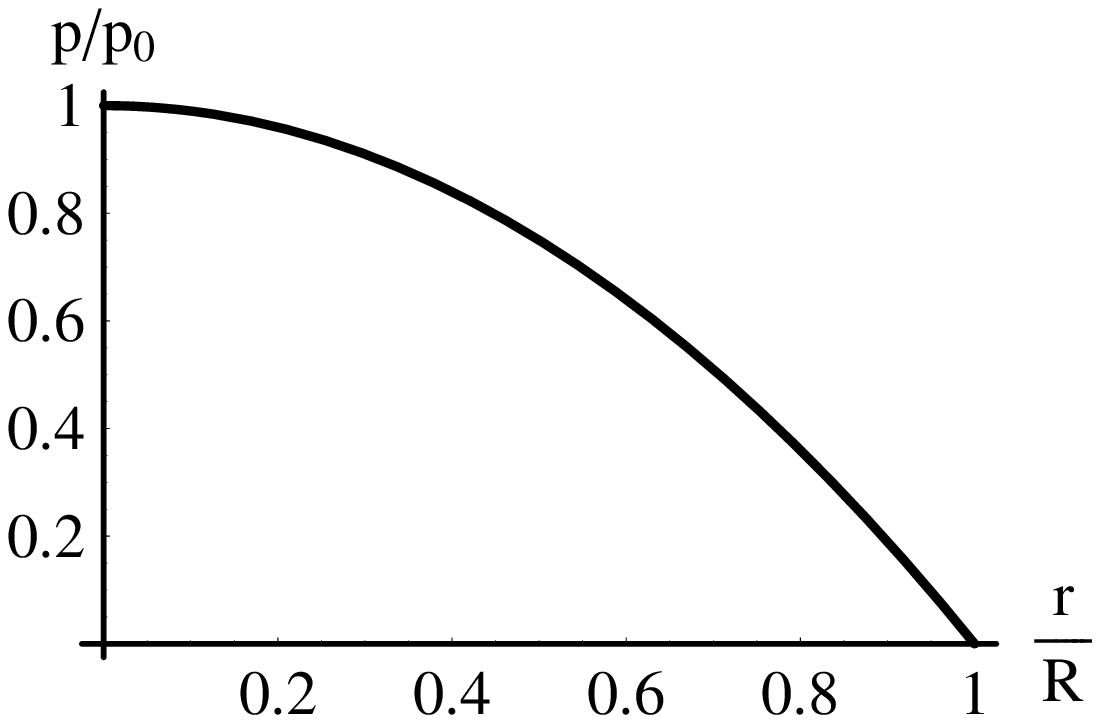}
\end{center}
\caption{Distribui\c{c}\~{o}es de $\protect\rho $, $M_{r}$, $p$ para $%
\protect\rho =const$. }
\label{f2.00}
\end{figure}

Escolhemos que a densidade da subst\^{a}ncia dentro da esfera seja constante%
\begin{equation}
\rho \left( r\right) =\rho _{0}=const~.  \label{2.08}
\end{equation}%
Pela Eq. (\ref{2.03}) temos para $M_{r}$:%
\begin{equation}
M_{r}=\int_{0}^{r}4\pi \rho _{0}r^{2}dr=\frac{4\pi }{3}\rho _{0}r^{3}=\frac{%
4\pi \rho _{0}R^{3}}{3}\left( \frac{r}{R}\right) ^{3}~,  \label{2.09}
\end{equation}%
sendo que a massa total da esfera \'{e}:%
\begin{equation}
M_{R}=M_{r}\left( R\right) =\frac{4\pi \rho _{0}}{3}R^{3}~.  \label{2.13}
\end{equation}%
Para determina\c{c}\~{a}o de $p\left( r\right) $ da Eq. (\ref{2.06}) temos a
equa\c{c}\~{a}o diferencial:%
\begin{equation*}
\frac{dp}{dr}=-G\frac{4\pi \rho _{0}}{3}R^{3}\left( \frac{r}{R}\right) ^{3}%
\frac{\rho _{0}}{r^{2}}
\end{equation*}%
ou%
\begin{equation}
\frac{dp}{dr}=-\frac{4\pi G\rho _{0}^{2}}{3}R\frac{r}{R}~,  \label{2.10}
\end{equation}%
com a condi\c{c}\~{a}o de contorno (\ref{2.07}). Resolvendo a Eq. (\ref{2.10}%
), temos:%
\begin{equation*}
p\left( r\right) -p_{0}=-\frac{4\pi G\rho _{0}^{2}}{3}R^{2}\int_{0}^{r}\frac{%
r}{R}d\left( \frac{r}{R}\right) =-\frac{4\pi G\rho _{0}^{2}}{3}R^{2}\frac{1}{%
2}\left( \frac{r}{R}\right) ^{2}~,
\end{equation*}%
onde $p_{0}=p\left( 0\right) $ \'{e} press\~{a}o no ponto $r=0$, que
apresenta a press\~{a}o cetral na esfera de g\'{a}s. Aplicando a condi\c{c}%
\~{a}o de contorno (\ref{2.07}), obtemos:%
\begin{equation}
p_{0}=\frac{1}{2}\frac{4\pi G\rho _{0}^{2}}{3}R^{2}~,  \label{2.11}
\end{equation}%
portanto:%
\begin{equation}
p\left( r\right) =\frac{4\pi G\rho _{0}^{2}}{3}\frac{1}{2}\left( 1-\frac{%
r^{2}}{R^{2}}\right) ~.  \label{2.12}
\end{equation}%
As distribui\c{c}\~{o}es de $\rho \left( r\right) $, $M_{r}\left( r\right) $%
, $p\left( r\right) $ s\~{a}o mostrados na Fig. \ref{f2.00}.

\subsection{Fun\c{c}\~{a}o de Distribui\c{c}\~{a}o da Densidade na Forma
Potencial}

\begin{figure}[th]
\includegraphics[width=9cm]{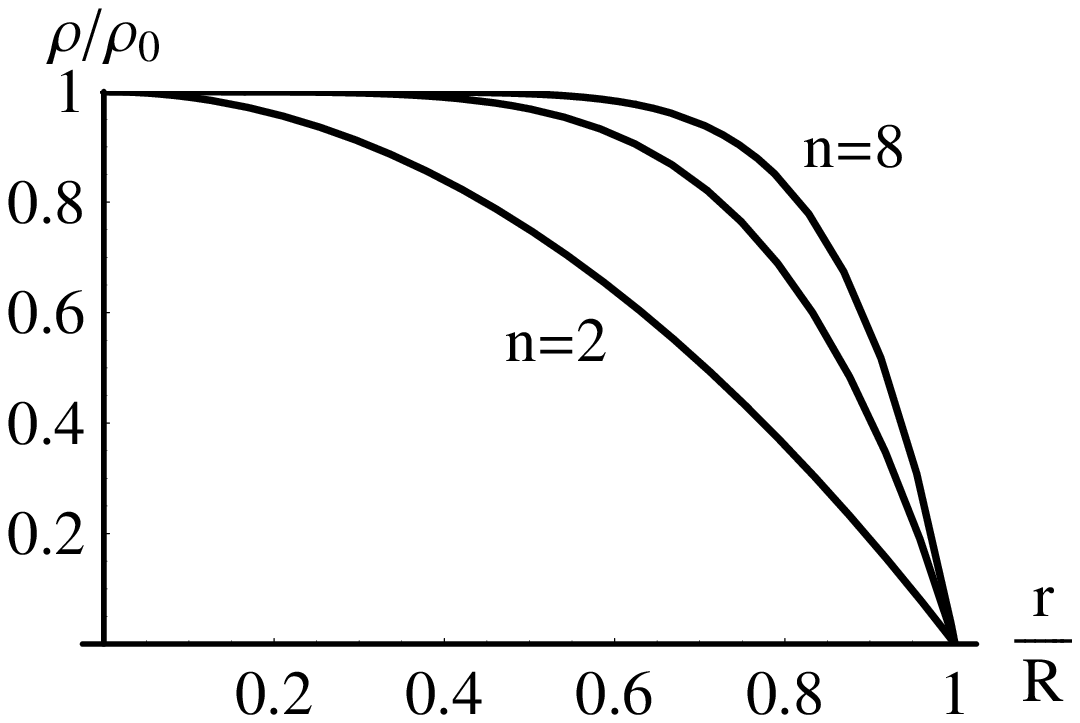} %
\includegraphics[width=9cm]{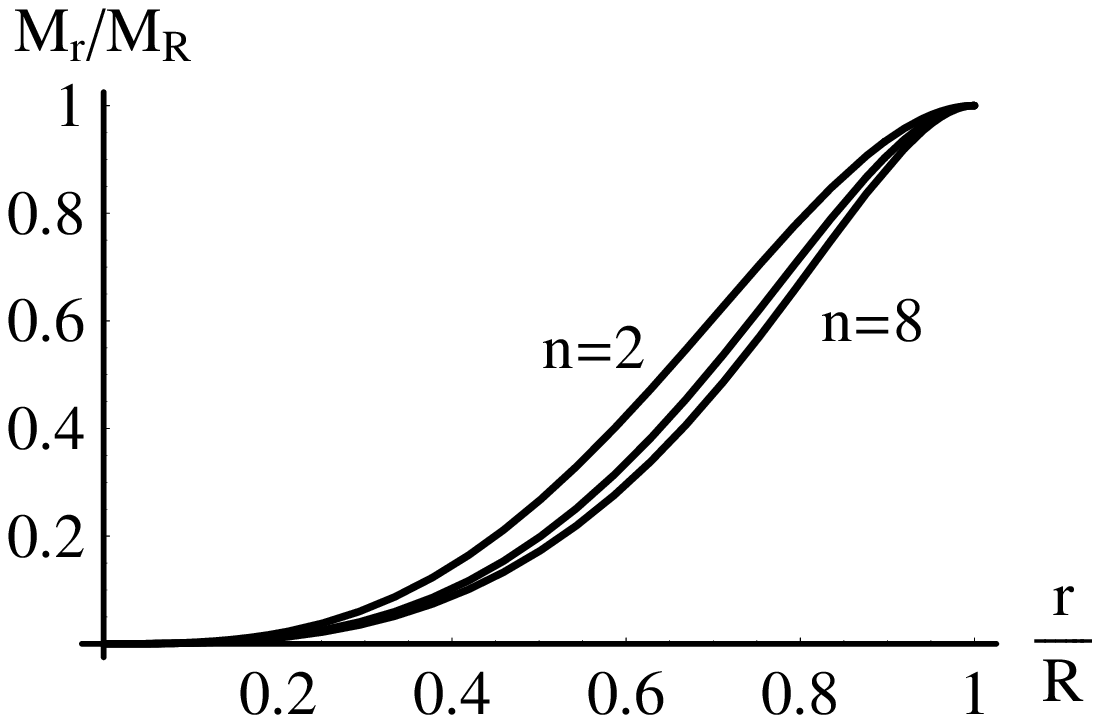}
\par
\begin{center}
\includegraphics[width=10cm]{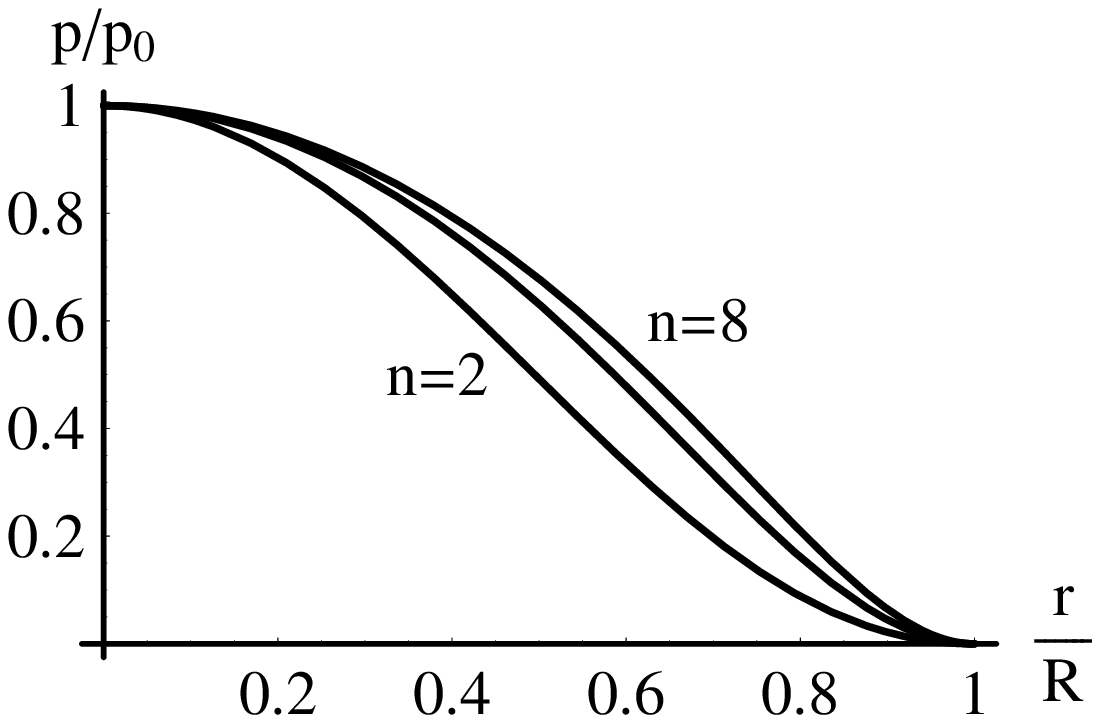}
\end{center}
\caption{Distribui\c{c}\~{o}es de $\protect\rho $, $M_{r}$, $p$ para $%
\protect\rho $ na forma potencial com $n=2,5,8$. Nos gr\'{a}ficos a curva
que corresponde a $n=5$ \'{e} intermei\'{a}ria entre as curvas de $n=2$ e $%
n=8$.}
\label{f2.04}
\end{figure}
Escolhemos a distribui\c{c}\~{a}o de densidade da subst\^{a}ncia na forma
potencial%
\begin{equation}
\rho \left( r\right) =\rho _{0}\left( 1-\frac{r^{n}}{R^{n}}\right) ~.
\label{2.14}
\end{equation}%
Pela Eq. (\ref{2.03}) temos para $M_{r}$:%
\begin{equation}
M_{r}=\int_{0}^{r}4\pi \rho _{0}\left( 1-\frac{r^{n}}{R^{n}}\right) r^{2}dr=%
\frac{4\pi \rho _{0}R^{3}}{3}\frac{r^{3}}{R^{3}}\left[ 1-\frac{3}{n+3}\frac{%
r^{n}}{R^{n}}\right] ~,  \label{2.15}
\end{equation}%
sendo que a massa total da esfera \'{e}:%
\begin{equation}
M_{R}=M_{r}\left( R\right) =\frac{4\pi \rho _{0}R^{3}}{3}\frac{n}{n+3}~.
\label{2.16}
\end{equation}%
Para determina\c{c}\~{a}o de $p\left( r\right) $ da Eq. (\ref{2.06}) temos a
equa\c{c}\~{a}o diferencial:%
\begin{equation}
\frac{dp}{dr}=-\frac{4\pi G\rho _{0}^{2}R}{3}\frac{r}{R}\left( 1-\frac{n+6}{%
n+3}\frac{r^{n}}{R^{n}}+\frac{3}{n+3}\frac{r^{2n}}{R^{2n}}\right) ~.
\label{2.17}
\end{equation}%
Resolvendo a Eq. (\ref{2.17}), temos:%
\begin{equation}
p\left( r\right) -p_{0}=-\frac{4\pi G\rho _{0}^{2}}{3}R^{2}\frac{1}{2}\frac{%
r^{2}}{R^{2}}\left[ 1-\frac{2\left( n+6\right) }{\left( n+3\right) \left(
n+2\right) }\frac{r^{n}}{R^{n}}+\frac{3}{\left( n+3\right) \left( n+1\right) 
}\frac{r^{2n}}{R^{2n}}\right] ~.  \label{2.18}
\end{equation}%
Aplicando a condi\c{c}\~{a}o de contorno (\ref{2.07}), obtemos:%
\begin{equation}
p_{0}=\frac{4\pi G\rho _{0}^{2}}{3}R^{2}\frac{1}{2}\left[ 1-\frac{2\left(
n+6\right) }{\left( n+3\right) \left( n+2\right) }+\frac{3}{\left(
n+3\right) \left( n+1\right) }\right] ~,  \label{2.19}
\end{equation}%
portanto%
\begin{equation*}
p\left( r\right) =\frac{4\pi G\rho _{0}^{2}}{3}R^{2}\frac{1}{2}\left\{
\left( 1-\frac{2\left( n+6\right) }{\left( n+3\right) \left( n+2\right) }+%
\frac{3}{\left( n+3\right) \left( n+1\right) }\right) \right.
\end{equation*}%
\begin{equation}
\left. -\frac{r^{2}}{R^{2}}\left( 1-\frac{2\left( n+6\right) }{\left(
n+3\right) \left( n+2\right) }\frac{r^{n}}{R^{n}}+\frac{3}{\left( n+3\right)
\left( n+1\right) }\frac{r^{2n}}{R^{2n}}\right) \right\} ~.  \label{2.20}
\end{equation}%
As distribui\c{c}\~{o}es de $\rho \left( r\right) $, $M_{r}\left( r\right) $%
, $p\left( r\right) $ s\~{a}o mostradas na Fig. \ref{f2.04} para $n=2,\ 5,\
8 $. No gr\'{a}fico as distribui\c{c}\~{o}es de $M_{r}\left( r\right) $ s%
\~{a}o fun\c{c}\~{o}es crescentes. Observamos tamb\'{e}m que na fun\c{c}\~{a}%
o de distribui\c{c}\~{a}o de densidade (\ref{2.14}) a pot\^{e}ncia $n$ n\~{a}%
o \'{e} necessariamente um n\'{u}mero inteiro e pode ser qualquer real $n>1$.

\subsection{Fun\c{c}\~{a}o de Distribui\c{c}\~{a}o da Densidade na Forma
Exponencial}

\begin{figure}[th]
\begin{center}
\includegraphics[width=12cm]{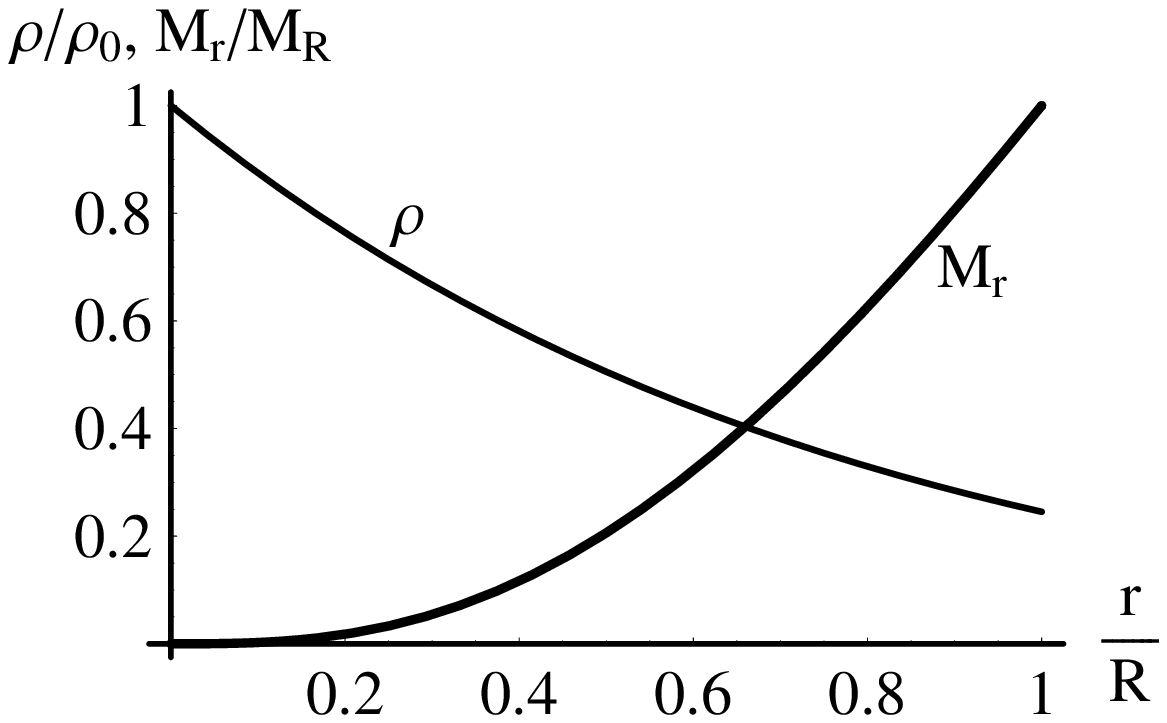} %
\includegraphics[width=12cm]{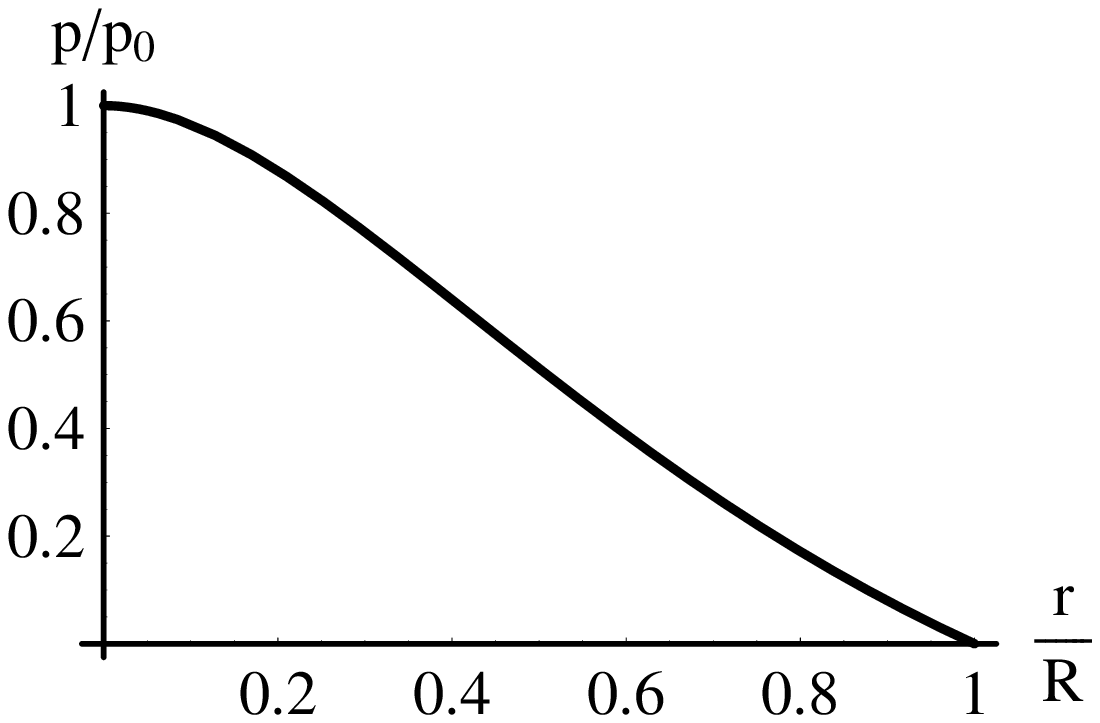}
\end{center}
\caption{Distribui\c{c}\~{o}es de $\protect\rho $, $M_{r}$, $p$ para $%
\protect\rho $ na forma exponencial. }
\label{f2.02}
\end{figure}
Escolhemos a distrubui\c{c}\~{a}o de densidade da subst\^{a}ncia na forma:%
\begin{equation}
\rho \left( r\right) =\rho _{0}\left[ 1-\frac{r}{3R}\right] e^{-\frac{r}{R}%
}~,  \label{2.21}
\end{equation}%
sendo que%
\begin{equation}
\rho \left( R\right) =\rho _{0}\left[ 1-\frac{1}{3}\right] \frac{1}{e}=\frac{%
2}{3e}\rho _{0}~.  \label{2.22}
\end{equation}%
Pela Eq. (\ref{2.03}) temos para $M_{r}$:%
\begin{equation*}
M_{r}=4\pi \rho _{0}\int_{0}^{r}\left[ 1-\frac{r}{3R}\right] e^{-\frac{r}{R}%
}r^{2}dr
\end{equation*}%
\begin{equation*}
=4\pi \rho _{0}R^{3}\int_{0}^{r}\left[ 1-\frac{r}{3R}\right] e^{-\frac{r}{R}%
}\left( \frac{r}{R}\right) ^{2}d\left( \frac{r}{R}\right)
\end{equation*}%
\begin{equation}
=4\pi \rho _{0}R^{3}\left[ \int_{0}^{r}e^{-\frac{r}{R}}\left( \frac{r}{R}%
\right) ^{2}d\left( \frac{r}{R}\right) -\frac{1}{3}\int_{0}^{r}e^{-\frac{r}{R%
}}\left( \frac{r}{R}\right) ^{3}d\left( \frac{r}{R}\right) \right] ~.
\label{2.23}
\end{equation}%
Na Eq. (\ref{2.23}), usando a t\'{e}cnica de integra\c{c}\~{a}o por partes:%
\begin{equation}
\int_{0}^{r}e^{-\frac{r}{R}}\left( \frac{r}{R}\right) ^{3}d\left( \frac{r}{R}%
\right) =-\left( \frac{r}{R}\right) ^{3}e^{-\frac{r}{R}}+3\int_{0}^{r}e^{-%
\frac{r}{R}}\left( \frac{r}{R}\right) ^{2}d\left( \frac{r}{R}\right) ~,
\label{2.24}
\end{equation}%
portanto:%
\begin{equation}
M_{r}=\frac{4\pi \rho _{0}}{3}R^{3}\left( \frac{r}{R}\right) ^{3}e^{-\frac{r%
}{R}}~,  \label{2.25}
\end{equation}%
sendo que a massa total da esfera \'{e}:%
\begin{equation}
M_{R}=M_{r}\left( R\right) =\frac{4\pi \rho _{0}}{3e}R^{3}~,  \label{2.26}
\end{equation}%
que permite tamb\'{e}m escrever $M_{r}$ como:%
\begin{equation}
M_{r}=M_{R}\left( \frac{r}{R}\right) ^{3}e^{1-\frac{r}{R}}~.  \label{2.27}
\end{equation}%
Para determina\c{c}\~{a}o de $p\left( r\right) $ da Eq. (\ref{2.06}) temos a
equa\c{c}\~{a}o diferencial:%
\begin{equation}
\frac{dp}{dr}=-\frac{4\pi G\rho _{0}^{2}R}{3}\frac{r}{R}\left( 1-\frac{1}{3}%
\frac{r}{R}\right) e^{-2\frac{r}{R}}~.  \label{2.28}
\end{equation}%
Resolvendo a Eq. (\ref{2.28}), temos:%
\begin{equation*}
p\left( r\right) -p_{0}=-\frac{4\pi G\rho _{0}^{2}R^{2}}{3}\int_{0}^{r}\frac{%
r}{R}\left( 1-\frac{1}{3}\frac{r}{R}\right) e^{-2\frac{r}{R}}d\left( \frac{r%
}{R}\right)
\end{equation*}%
\begin{equation}
=-\frac{4\pi G\rho _{0}^{2}R^{2}}{3}\left[ \int_{0}^{r}\left( \frac{r}{R}%
\right) e^{-2\frac{r}{R}}d\left( \frac{r}{R}\right) -\frac{1}{3}%
\int_{0}^{r}\left( \frac{r}{R}\right) ^{2}e^{-2\frac{r}{R}}d\left( \frac{r}{R%
}\right) \right] ~.  \label{2.29}
\end{equation}%
Na Eq. (\ref{2.29}):%
\begin{equation}
\int_{0}^{r}\left( \frac{r}{R}\right) ^{2}e^{-2\frac{r}{R}}d\left( \frac{r}{R%
}\right) =-\frac{1}{2}\left( \frac{r}{R}\right) ^{2}e^{-2\frac{r}{R}%
}+\int_{0}^{r}\left( \frac{r}{R}\right) e^{-2\frac{r}{R}}d\left( \frac{r}{R}%
\right) ~,  \label{2.30}
\end{equation}%
portanto:%
\begin{equation}
p\left( r\right) -p_{0}=-\frac{4\pi G\rho _{0}^{2}R^{2}}{3}\left[ \frac{2}{3}%
\int_{0}^{r}\left( \frac{r}{R}\right) e^{-2\frac{r}{R}}d\left( \frac{r}{R}%
\right) +\frac{1}{6}\left( \frac{r}{R}\right) ^{2}e^{-2\frac{r}{R}}\right] ~.
\label{2.31}
\end{equation}%
A integral%
\begin{equation}
\int_{0}^{r}\left( \frac{r}{R}\right) e^{-2\frac{r}{R}}d\left( \frac{r}{R}%
\right) =-\frac{1}{2}e^{-2\frac{r}{R}}\left( \frac{r}{R}+\frac{1}{2}\right) +%
\frac{1}{4}~,  \label{2.32}
\end{equation}%
portanto%
\begin{equation}
p\left( r\right) -p_{0}=-\frac{4\pi G\rho _{0}^{2}R^{2}}{3}\left[ \frac{1}{6}%
+\left( -\frac{1}{6}-\frac{1}{3}\frac{r}{R}+\frac{1}{6}\left( \frac{r}{R}%
\right) ^{2}\right) e^{-2\frac{r}{R}}\right] ~.  \label{2.33}
\end{equation}%
Aplicando a condi\c{c}\~{a}o de contorno (\ref{2.07}), obtemos:%
\begin{equation}
p_{0}=G\frac{4\pi \rho _{0}^{2}R^{2}}{3}\left( \frac{1}{6}-\frac{1}{3e^{2}}%
\right) ~,  \label{2.34}
\end{equation}%
portanto%
\begin{equation*}
p\left( r\right) =\frac{4\pi G\rho _{0}^{2}R^{2}}{3}\left[ \left( \frac{1}{6}%
-\frac{1}{3e^{2}}\right) -\left( \frac{1}{6}+\left( -\frac{1}{6}-\frac{1}{3}%
\frac{r}{R}+\frac{1}{6}\left( \frac{r}{R}\right) ^{2}\right) e^{-2\frac{r}{R}%
}\right) \right]
\end{equation*}%
\begin{equation}
=\frac{4\pi G\rho _{0}^{2}R^{2}}{3}\frac{1}{3e^{2}}\left[ \left( \frac{1}{2}+%
\frac{r}{R}-\frac{1}{2}\left( \frac{r}{R}\right) ^{2}\right) e^{2\left( 1-%
\frac{r}{R}\right) }-1\right] ~.  \label{2.35}
\end{equation}%
As distribui\c{c}\~{o}es de $\rho \left( r\right) $, $M_{r}\left( r\right) $%
, $p\left( r\right) $ s\~{a}o mostrados na Fig. \ref{f2.02}.

\subsection{Fun\c{c}\~{a}o de Distribui\c{c}\~{a}o da Densidade na Forma
Gaussiana}

\begin{figure}[th]
\begin{center}
\includegraphics[width=12cm]{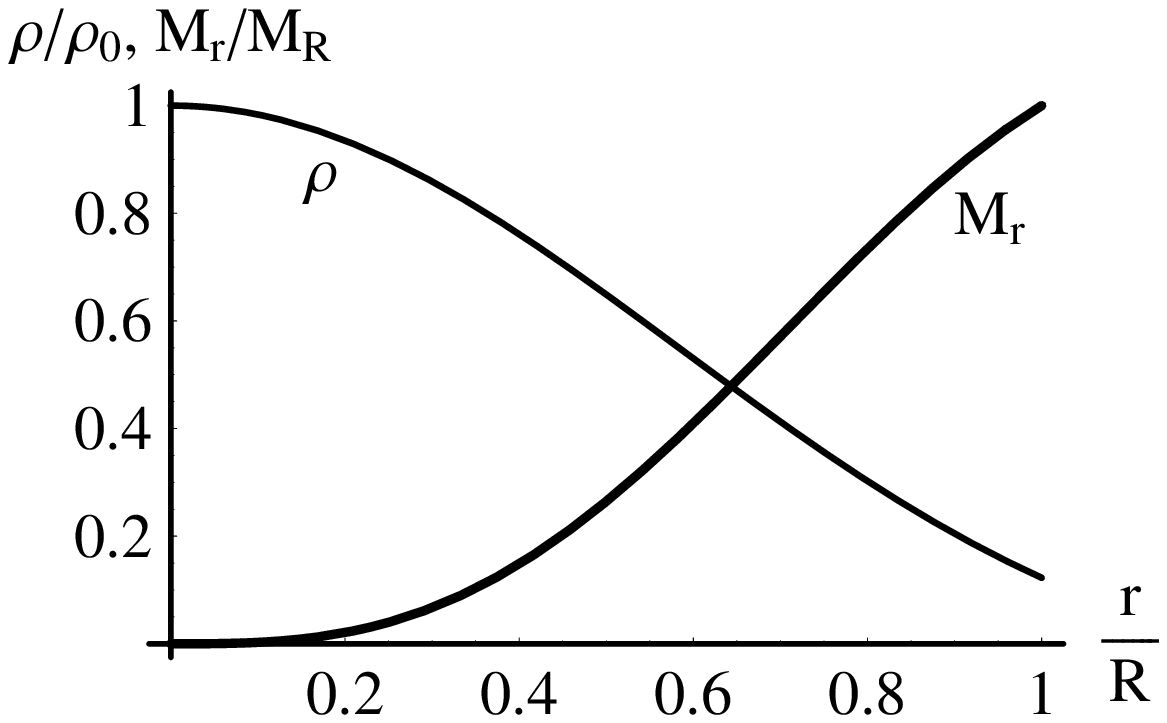} %
\includegraphics[width=12cm]{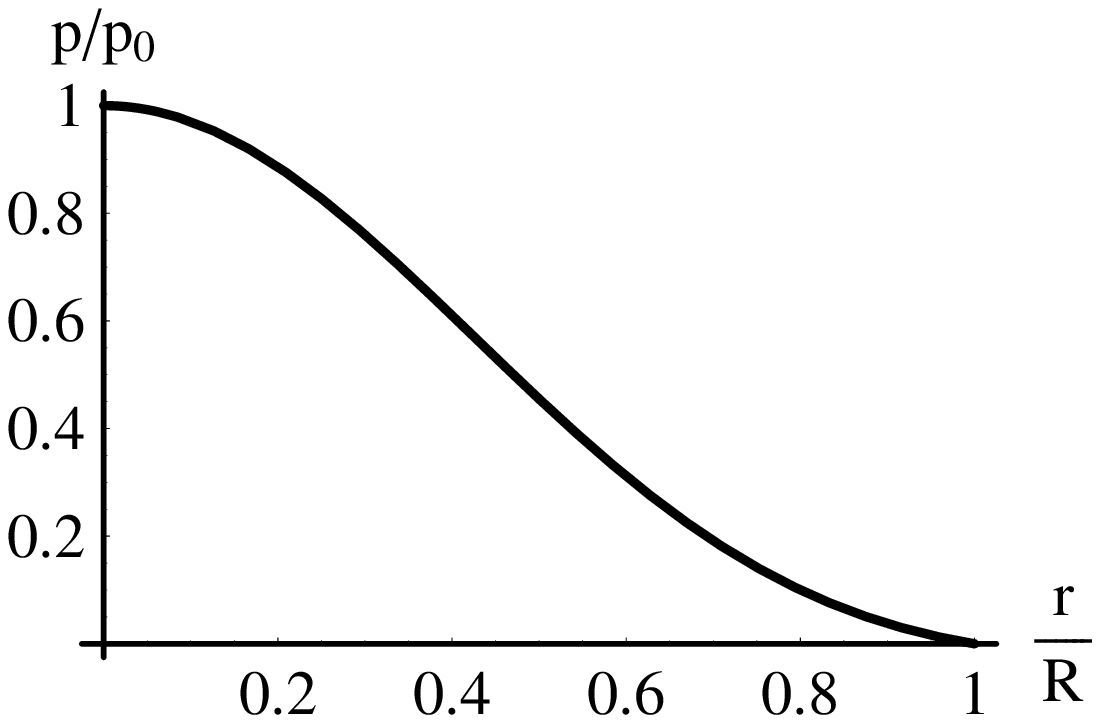}
\end{center}
\caption{Distribui\c{c}\~{o}es de $\protect\rho $, $M_{r}$, $p$ para $%
\protect\rho $ na forma gaussiana. }
\label{f2.03}
\end{figure}
Escolhemos a distrubui\c{c}\~{a}o de densidade da subst\^{a}ncia na forma
gaussiana:%
\begin{equation}
\rho \left( r\right) =\rho _{0}\left[ 1-\frac{2}{3}\frac{r^{2}}{R^{2}}\right]
e^{-\frac{r^{2}}{R^{2}}}~,  \label{2.36}
\end{equation}%
sendo que:%
\begin{equation}
\rho \left( R\right) =\rho _{0}\left[ 1-\frac{2}{3}\right] \frac{1}{e}=\frac{%
1}{3e}\rho _{0}\ .  \label{2.37}
\end{equation}%
Pela Eq. (\ref{2.03}) temos para $M_{r}$:%
\begin{equation*}
M_{r}=4\pi \rho _{0}\int_{0}^{r}\left[ 1-\frac{2}{3}\frac{r^{2}}{R^{2}}%
\right] e^{-\frac{r^{2}}{R^{2}}}r^{2}dr
\end{equation*}%
\begin{equation*}
=4\pi \rho _{0}R^{3}\int_{0}^{r}\left[ 1-\frac{2}{3}\frac{r^{2}}{R^{2}}%
\right] e^{-\frac{r^{2}}{R^{2}}}\left( \frac{r}{R}\right) ^{2}d\left( \frac{r%
}{R}\right)
\end{equation*}%
\begin{equation}
=4\pi \rho _{0}R^{3}\left[ \int_{0}^{r}e^{-\frac{r^{2}}{R^{2}}}\left( \frac{r%
}{R}\right) ^{2}d\left( \frac{r}{R}\right) -\frac{2}{3}\int_{0}^{r}e^{-\frac{%
r^{2}}{R^{2}}}\left( \frac{r}{R}\right) ^{4}d\left( \frac{r}{R}\right) %
\right] ~.  \label{2.38}
\end{equation}%
Observando que:%
\begin{equation}
de^{-x^{2}}=-2xe^{-x^{2}}dx~,  \label{2.39}
\end{equation}%
temos para a integral:%
\begin{equation*}
\int_{0}^{r}e^{-\frac{r^{2}}{R^{2}}}\left( \frac{r}{R}\right) ^{4}d\left( 
\frac{r}{R}\right) =-\frac{1}{2}\int_{0}^{r}\left( \frac{r}{R}\right)
^{3}de^{-\frac{r^{2}}{R^{2}}}=-\frac{1}{2}\left[ \left. \left( \frac{r}{R}%
\right) ^{3}e^{-\frac{r^{2}}{R^{2}}}\right\vert _{0}^{r}-\int_{0}^{r}e^{-%
\frac{r^{2}}{R^{2}}}d\left( \frac{r}{R}\right) ^{3}\right]
\end{equation*}%
\begin{equation}
=-\frac{1}{2}\left[ \left( \frac{r}{R}\right) ^{3}e^{-\frac{r^{2}}{R^{2}}%
}-3\int_{0}^{r}e^{-\frac{r^{2}}{R^{2}}}\left( \frac{r}{R}\right) ^{2}d\left( 
\frac{r}{R}\right) \right] ~,  \label{2.40}
\end{equation}%
que d\'{a}:%
\begin{equation}
M_{r}=4\pi \rho _{0}R^{3}\left[ \int_{0}^{r}e^{-\frac{r^{2}}{R^{2}}}\left( 
\frac{r}{R}\right) ^{2}d\left( \frac{r}{R}\right) +\frac{1}{3}\left( \frac{r%
}{R}\right) ^{3}e^{-\frac{r^{2}}{R^{2}}}-\int_{0}^{r}e^{-\frac{r^{2}}{R^{2}}%
}\left( \frac{r}{R}\right) ^{2}d\left( \frac{r}{R}\right) \right] ~.
\label{2.41}
\end{equation}%
Portanto:%
\begin{equation}
M_{r}=\frac{4\pi \rho _{0}R^{3}}{3}\left( \frac{r}{R}\right) ^{3}e^{-\frac{%
r^{2}}{R^{2}}}~,  \label{2.42}
\end{equation}%
sendo que a massa total da esfera \'{e}:%
\begin{equation}
M_{R}=M_{r}\left( R\right) =\frac{4\pi \rho _{0}}{3e}R^{3}~,  \label{2.43}
\end{equation}%
que permite tamb\'{e}m escrever $M_{r}$ como:%
\begin{equation}
M_{r}=M_{R}\left( \frac{r}{R}\right) ^{3}e^{1-\frac{r^{2}}{R^{2}}}~.
\label{2.44}
\end{equation}%
Para determina\c{c}\~{a}o de $p\left( r\right) $ da Eq. (\ref{2.06}) temos a
equa\c{c}\~{a}o diferencial:%
\begin{equation}
\frac{dp}{dr}=-\frac{4\pi G\rho _{0}^{2}}{3}R\frac{r}{R}\left[ 1-\frac{2}{3}%
\frac{r^{2}}{R^{2}}\right] e^{-2\frac{r^{2}}{R^{2}}}~.  \label{2.45}
\end{equation}%
Resolvendo a Eq. (\ref{2.45}), temos:%
\begin{equation*}
p\left( r\right) -p_{0}=-\frac{4\pi G\rho _{0}^{2}R^{2}}{3}\int_{0}^{r}\frac{%
r}{R}\left[ 1-\frac{2}{3}\frac{r^{2}}{R^{2}}\right] e^{-2\frac{r^{2}}{R^{2}}%
}d\frac{r}{R}
\end{equation*}%
\begin{equation}
=-\frac{4\pi G\rho _{0}^{2}R^{2}}{3}\left[ \int_{0}^{r}\frac{r}{R}e^{-2\frac{%
r^{2}}{R^{2}}}d\frac{r}{R}-\frac{2}{3}\int_{0}^{r}\frac{r^{3}}{R^{3}}e^{-2%
\frac{r^{2}}{R^{2}}}d\frac{r}{R}\right] ~.  \label{2.46}
\end{equation}%
Na Eq. (\ref{2.46}) a integral:%
\begin{equation*}
\int_{0}^{r}\frac{r^{3}}{R^{3}}e^{-2\frac{r^{2}}{R^{2}}}d\frac{r}{R}=-\frac{1%
}{4}\left[ \int_{0}^{r}\frac{r^{2}}{R^{2}}de^{-2\frac{r^{2}}{R^{2}}}\right]
=-\frac{1}{4}\left[ \frac{r^{2}}{R^{2}}e^{-2\frac{r^{2}}{R^{2}}%
}-\int_{0}^{r}e^{-2\frac{r^{2}}{R^{2}}}d\frac{r^{2}}{R^{2}}\right]
\end{equation*}%
\begin{equation}
=-\frac{1}{4}\left[ \frac{r^{2}}{R^{2}}e^{-2\frac{r^{2}}{R^{2}}}+\frac{1}{2}%
\left( e^{-2\frac{r^{2}}{R^{2}}}-1\right) \right] =-\frac{1}{4}\left[ \left( 
\frac{r^{2}}{R^{2}}+\frac{1}{2}\right) e^{-2\frac{r^{2}}{R^{2}}}-\frac{1}{2}%
\right]  \label{2.47}
\end{equation}%
e a integral%
\begin{equation}
\int_{0}^{r}\frac{r}{R}e^{-2\frac{r^{2}}{R^{2}}}d\frac{r}{R}=\frac{1}{2}%
\int_{0}^{r}e^{-2\frac{r^{2}}{R^{2}}}d\left( \frac{r^{2}}{R^{2}}\right) =-%
\frac{1}{4}\int_{0}^{r}e^{-2\frac{r^{2}}{R^{2}}}d\left( \frac{-2r^{2}}{R^{2}}%
\right) =-\frac{1}{4}\left( e^{-2\frac{r^{2}}{R^{2}}}-1\right) ~.
\label{2.48}
\end{equation}%
Portanto:%
\begin{equation*}
p\left( r\right) -p_{0}=-\frac{4\pi G\rho _{0}^{2}R^{2}}{3}\left[ -\frac{1}{4%
}\left( e^{-2\frac{r^{2}}{R^{2}}}-1\right) +\frac{1}{6}\left[ \left( \frac{%
r^{2}}{R^{2}}+\frac{1}{2}\right) e^{-2\frac{r^{2}}{R^{2}}}-\frac{1}{2}\right]
\right]
\end{equation*}%
\begin{equation}
=-G\frac{4\pi \rho _{0}^{2}R^{2}}{3}\frac{1}{6}\left[ 1+\left( \frac{r^{2}}{%
R^{2}}-1\right) e^{-2\frac{r^{2}}{R^{2}}}\right] \ .  \label{2.49}
\end{equation}%
Aplicando a condi\c{c}\~{a}o de contorno (\ref{2.07}), obtemos:%
\begin{equation}
p_{0}=G\frac{4\pi \rho _{0}^{2}R^{2}}{3}\frac{1}{6}\ .  \label{2.50}
\end{equation}%
Portanto%
\begin{equation}
p\left( r\right) =\frac{4\pi G\rho _{0}^{2}R^{2}}{3}\frac{1}{6}\left( 1-%
\frac{r^{2}}{R^{2}}\right) e^{-2\frac{r^{2}}{R^{2}}}\ .  \label{2.51}
\end{equation}%
As distribui\c{c}\~{o}es de $\rho \left( r\right) $, $M_{r}\left( r\right) $%
, $p\left( r\right) $ s\~{a}o mostrados na Fig. \ref{f2.03}.

\section{Conclus\~{a}o}

Neste trabalho foram discutidos modelos estrelares anal\'{\i}ticos
simplificados que incluem duas equa\c{c}\~{o}es b\'{a}sicas: a equa\c{c}\~{a}%
o de equil\'{\i}brio hidrost\'{a}tico e a equa\c{c}\~{a}o de conserva\c{c}%
\~{a}o de massa. Usando distribui\c{c}\~{a}o de densidade da massa na
estrela como uma determinada fun\c{c}\~{a}o obtemos analiticamente a
correspondente distribui\c{c}\~{a}o de press\~{a}o dentro da esfera e
determinamos a press\~{a}o central. Discutimos as distribui\c{c}\~{o}es de
densidade da forma constante, potencial, exponencial e gaussiana. As
distribui\c{c}\~{o}es propostas no trabalho s\~{a}o at\'{e} um determinado
grau convencionais. Entretanto elas s\~{a}o mais real\'{\i}sticas (exceto o
exemplo de densidade constante) comparando, por exemplo, com o modelo
estrelar linear.

Para modelagem de estrelas com um n\'{u}cleo extenso \'{e} mais apropriada a
distribui\c{c}\~{a}o da forma potencial; Varia\c{c}\~{a}o do grau da fun\c{c}%
\~{a}o potencial possibilita modelar a extens\~{a}o do nucleo estrelar. Para
modelagem de alguns tipos de estrelas \'{e} mais apropriado utilizar um
modelo com densidade na superf\'{\i}cie n\~{a}o nula. As distribui\c{c}\~{o}%
es exponencial e gaussiana realizam esse mesmo cen\'{a}rio. Em alguns casos
pode ser conveniente fazer estima\c{c}\~{o}es, utilizando a condi\c{c}\~{a}o
de contorno no infinito. Os modelos com as distribui\c{c}\~{o}es exponencial
e gaussiana podem ser facilmente reformulados para a condi\c{c}\~{a}o de
contorno no infinito: $p\left( r\right) =0$ quando $r\rightarrow \infty $.
No caso da distribui\c{c}\~{a}o exponencial a correspondente solu\c{c}\~{a}o
toma \`{a} forma:%
\begin{equation}
p\left( r\right) =-\frac{4\pi G\rho _{0}^{2}R^{2}}{3}\frac{1}{6}\left( 1+2%
\frac{r}{R}-\left( \frac{r}{R}\right) ^{2}\right) e^{-2\frac{r}{R}}~,
\label{2.52}
\end{equation}%
com a press\~{a}o central 
\begin{equation}
p_{0}=\frac{4\pi G\rho _{0}^{2}R^{2}}{3}\frac{1}{6}~.  \label{2.53}
\end{equation}%
No caso da distribui\c{c}\~{a}o gaussiana a solu\c{c}\~{a}o tem a mesma
forma das Eqs. (\ref{2.50}), (\ref{2.51})%
\begin{equation}
p\left( r\right) =\frac{4\pi G\rho _{0}^{2}R^{2}}{3}\frac{1}{6}\left( 1-%
\frac{r^{2}}{R^{2}}\right) e^{-2\frac{r^{2}}{R^{2}}},\ p_{0}=G\frac{4\pi
\rho _{0}^{2}R^{2}}{3}\frac{1}{6}~.  \label{2.54}
\end{equation}

\begin{figure}[th]
\begin{center}
\includegraphics[width=12cm]{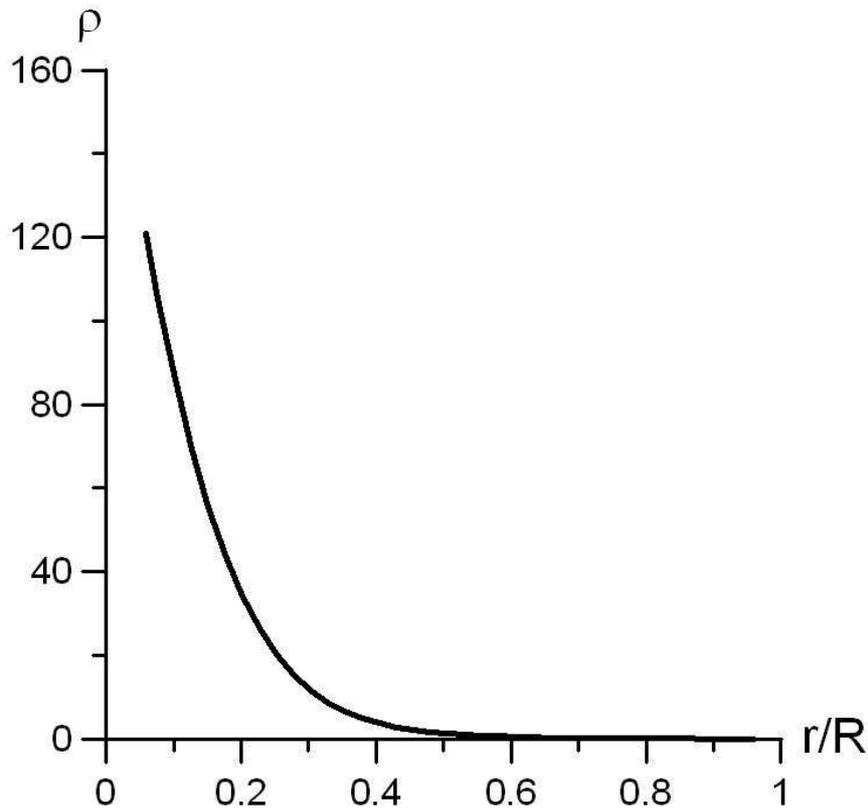}
\end{center}
\caption{Distribu\c{c}\~{a}o da densidade de massa no Sol segundo os dados
de BiSON-13 \protect\cite{Basu}; a densidade $\protect\rho $ \'{e}
apresentada em g/cm$^{3}$.}
\label{f-c01}
\end{figure}

Para dar um exemplo de uma distribui\c{c}\~{a}o da densidade dentro de
estrelas reais apresentamos na figura (\ref{f-c01}) a distribui\c{c}\~{a}o
de densidade dentro do Sol obtida pelo Birmingham Solar Oscillations Network
(BiSON-13) da Ref. \cite{Basu}. Prestamos aten\c{c}\~{a}o que os dados na
Ref. \cite{Basu} iniciam do valor de $r/R_{\odot }=$ 0,058884 ($R_{\odot }$ 
\'{e} o raio do Sol). Portanto para construir o modelo de qualquer forma 
\'{e} necess\'{a}rio aplicar uma hip\'{o}tese sobre o intervalo central $%
0<r<r_{in}$. Para estrelas distantes \'{e} imposs\'{\i}vel obter a informa%
\c{c}\~{a}o tanta detalhada. Usando os dados observacionais diretos e
indiretos, fazem conclus\~{o}es sobre as caracter\'{\i}sticas mais gerais:
luminosidade, massa, raio, caracter\'{\i}sticas espectrais, etc.. Ent\~{a}o,
modelagem f\'{\i}sica matem\'{a}tica torna-se o \'{u}nico m\'{e}todo de
investiga\c{c}\~{a}o da estrutura interna das estrelas.

As fun\c{c}\~{o}es propostas neste trabalho a t\'{\i}tulo de distribui\c{c}%
\~{o}es da densidade podem ser consideradas como fun\c{c}\~{o}es\ de teste.
Introdu\c{c}\~{a}o de um par\^{a}metro (ou v\'{a}rios par\^{a}metros) na fun%
\c{c}\~{a}o possibilita ajustar a fun\c{c}\~{a}o a um aspecto mais
apropriado pela varia\c{c}\~{a}o do par\^{a}metro. Se ainda satisfazer a
propriedade de obten\c{c}\~{a}o de solu\c{c}\~{o}es das equa\c{c}\~{o}es do
modelo na forma anal\'{\i}tica podemos chegar a um modelo (de fato a um
classe de modelos) anal\'{\i}tico exatamante sol\'{u}vel com possibilidade
de escolha e ajuste da fun\c{c}\~{a}o de distribui\c{c}\~{a}o da densidade.
Tanta discuss\~{a}o de modelos anal\'{\i}ticos exatamante sol\'{u}veis vemos
como uma das poss\'{\i}veis continua\c{c}\~{o}es deste trabalho.


\begin{thebibliography}{9}
\bibitem{Clayton} Clayton D. D., Principles of Stellar Evolution and
Nucleosynthesis (McGraw-Hill, New York) 1968.

\bibitem{Hansen} Hansen C. J., Kawaler S. D., Trimble V., Stellar Interiors
(Springer) 2004.

\bibitem{stein} Stein R. F., Stellar Evolution: a Survey with Analitic
Models, in Stellar Evolution, eds. Stein \& Cameron (NY: Plenum Press)
1966, pp. 3-82.

\bibitem{boas} Boas M. L., Mathematical Methods in the Physical Sciences
(John Wiley, NY) 2005.

\bibitem{ll-t2} Landau L. D., Lifshitz E. M., Course of Theoretical Physics,
vol. 2, The Classical Theory of Fields (Pergamon Press) 1971.

\bibitem{schutz} Schutz B. F., A First Course in General Relativity
(Cambridge University Press) 2009.

\bibitem{Basu} Basu et al., Fresh insights on the structure of the solar
core, The Astrophysical Journal, 699, 1403-1417 (2009).
\end{thebibliography}
\end{document}